\documentclass[final,3p,times]{elsarticle}
\usepackage{amssymb}
\usepackage{amsthm}
\usepackage{amsfonts}
\usepackage[latin1]{inputenc}

\usepackage{natbib}
\biboptions{authoryear}

\journal{Nucleation Theory and Applications 2011}

\begin{document}

\begin{frontmatter}

\title{Gas Bubble Growth Dynamics in a Supersaturated Solution: \\ Henry's and Sievert's Solubility Laws}

\author[label1]{Gennady Yu. Gor}
\ead{ggor@princeton.edu}
\author[label1]{Anatoly E. Kuchma}
\author[label1]{Fedor M. Kuni}

\address[label1]{Department of Physics, St. Petersburg State University \\ Ulyanovskaya Street 1, Petrodvorets, St. Petersburg, 198504, Russia}

\fnref{label2}
\fntext[label2]{Corresponding author. Current address: Department of Civil and Environmental Engineering, Princeton University, Princeton, New Jersey 08544, United States}

\begin{abstract}
Theoretical description of diffusion growth of a gas bubble after its nucleation in supersaturated liquid solution is presented. We study the influence of Laplace pressure on the bubble growth. We consider two different solubility laws: Henry's law, which is fulfilled for the systems where no gas molecules dissociation takes place and Sievert's law, which is fulfilled for the systems where gas molecules completely dissociate in the solvent into two parts. We show that the difference between Henry's and Sievert's laws for chemical equilibrium conditions causes the difference in bubble growth dynamics. Assuming that diffusion flux of dissolved gas molecules to the bubble is steady we obtain differential equations on bubble radius for both solubility laws. For the case of homogeneous nucleation of a bubble, which takes place at a significant pressure drop bubble dynamics equations for Henry's and Sievert's laws are solved analytically. For both solubility laws three characteristic stages of bubble growth are marked out. Intervals of bubble size change and time intervals of these stages are found. We also obtain conditions of diffusion flux steadiness corresponding to consecutive stages. The fulfillment of these conditions is discussed for the case of nucleation of water vapor bubbles in magmatic melts. For Sievert's law the analytical treatment of the problem of bubble dissolution in a pure solvent is also presented.

\end{abstract}

\end{frontmatter}

\section{Introduction}

This paper presents a theoretical description of diffusion growth of a gas bubble in liquid solution as a result of a considerable pressure drop (in the order of $ 10^3$ times). These conditions of bubble growth process are observed in magmatic melts during volcanic eruptions [\ref{Sparks-1978}, \ref{Sahagian-1999}]. After such a significant pressure drop the solution becomes strongly supersaturated; and homogeneous (fluctuational) nucleation of gas bubbles becomes possible. It is the growth dynamics of such bubbles that is the subject of the present paper. It has to be noted that the growth regularities of a solitary bubble are crucial for the description of the whole kinetics of phase transition in supersaturated solution [\ref{Chernov-2004}, \ref{Slezov-2004}, \ref{Slezov-2005}].

While describing gas bubble growth in supersaturated solution two rough approximations are traditionally made [\ref{Chernov-2004}, \ref{Navon}, \ref{Lensky}, \ref{Slezov-2004}, \ref{Slezov-2005}]:
\begin{enumerate}
\item The flux of the dissolved gas towards the bubble is assumed to be steady.
\item The consideration is limited to bubbles with the radius that is large enough to neglect the Laplace pressure in comparison with the external pressure of the solution.
\end{enumerate}
In the present paper, following our preceding works [\ref{Kuchma-Gor-Kuni}, \ref{Gor-Kuchma-Sievert}], while exploiting the steady approximation, we take into account the Laplace pressure in the bubble considering the bubble from the very moment of its nucleation. Consideration of the time-dependent Laplace pressure in the bubble makes both gas density in the bubble and the equilibrium concentration of the dissolved gas at the surface of the bubble time-dependent as well.

Gas bubble growth in a solution taking into account the Laplace pressure was considered as early as in 1950 in the classical paper by Epstein and Plesset [\ref{Epstein-Plesset}], where the authors obtained the equation for the bubble radius as a function of time. In order to relate the equilibrium concentration of the dissolved gas and the solution pressure, [\ref{Epstein-Plesset}] presupposed the fulfillment of Henry's law: i. e. the proportional dependence between these two values. Indeed, Henry's law is fulfilled for the solution of $\rm CO_2$ in magmatic melt [\ref{Lensky}]; however, it is not valid for the solution of $\rm H_2O$ vapor [\ref{Chernov-2004}, \ref{Navon}]. In this case, which is crucial for practical reasons, Sievert's law is observed: the equilibrium concentration of the dissolved gas is proportional to the square root of the solution pressure [\ref{Stolper}]. Sometimes both cases are referred to as "Henry's law", but in the present paper, in order to avoid confusion, we will use the term "Sievert's law" for the case with a square root, following e. g. [\ref{Cable-Frade}].
Here we analyze the bubble growth dynamics in both cases: for Henry's and Sievert's laws. For Sievert's law we obtain the equation for the bubble radius as a function of time analogous to [\ref{Epstein-Plesset}], which has not been obtained previously.

Our analysis shows that, irrespective of the law applied to gas solubility, three characteristic stages can be marked out in the growth dynamics. During the first stage the bubble radius is growing with an increasing rate. On the second stage the growth rate decreases. The third stage, when the growth rate continues to decrease, begins when the Laplace pressure inside the bubble becomes comparable with the external pressure of the solution. We demonstrate that during the first two stages the time dependence of the bubble radius is different for the cases when either  Henry's and Sievert's laws are fulfilled, while during the third stage this distinction is no longer observed.

For both Henry's and Sievert's laws we obtain intervals within which the bubble radius changes on each stage, as well as time limits and conditions when the steady approximation is applicable. We show that, as the radius of the bubble increases, the steady condition becomes stricter; and, consequently, as a rule, the steady regime of a multistage bubble growth gradually gives way to the nonsteady one.
We obtain analytical expression for the time when Laplace pressure influences on bubble growth vanishes and therefore substantiate the estimation of this time made in [\ref{Grinin-Kuni-Gor}]. After this time passed the bubble growth reaches a self-similar regime [\ref{Scriven-1959}].
We also analyze whether the steady approximation is applicable to the case of gas bubbles in magma described by Navon [\ref{Navon}] and Chernov et al. [\ref{Chernov-2004}] for large radius of a bubble (neglecting Laplace pressure).
Besides that, we present the analytical description of bubble dissolution in the pure solvent for the Sievert's solubility law, which was not presented before in literature.

\section{Equilibrium Concentration of the Dissolved Gas}

Let us consider a gas dissolved in a liquid. The solution was initially in an equilibrium state at temperature $T$ and pressure $P_0$. The concentration of the dissolved gas in the solution under such conditions will be denoted as $n_0$.
Then we instantly relieve the external pressure to value $\Pi$ in such way that solution becomes supersaturated. The temperature and volume of the solution remain the same, thus value $n_0$ still serves for the dissolved gas concentration.

It is more convenient to express the state of the solution in terms of dimensionless variables: supersaturation $\zeta$ and gas solubility $s$ defined here via
\begin{equation}
\label{zeta}
\zeta \equiv \frac{n_0 - n_\infty}{n_\infty}\; ,
\end{equation}
\begin{equation}
\label{s}
s \equiv \frac{k T n_\infty}{\Pi}\; ,
\end{equation}
where $n_{\infty}$ is the equilibrium concentration of dissolved gas at the external pressure $\Pi$, $k$ is the Boltzmann's constant.

When some time passes after the pressure drop, a gas bubble nucleates and begins growing regularly. Following [\ref{Chernov-2004}] we assume that the bubble is in mechanical equilibrium with the solution, and its dynamics is governed only by diffusion process. This assumption will be discussed in Appendix A. The radius of the bubble will be denoted as $R$. We consider the situation when the solvent is in its stable liquid state; therefore, we assume that the bubble consists of gas only, but not of the solvent vapor. The gas in the bubble is considered to be ideal.

When the bubble is studied after some time $t_0$ since its nucleation, its radius complies with the strong inequality
\begin{equation}
\label{R>>Laplace}
R \gg 2 \sigma /\Pi\; ,
\end{equation}
where $\sigma$ is the surface tension of the pure solvent (it is true while the solution is considered as diluted). Eq.~(\ref{R>>Laplace}) allows us to neglect the influence of Laplace pressure on the bubble growth. Therefore, we can write the following equations for the pressure in the bubble $P_R$ and for the equilibrium solution concentration near the surface of the bubble $n_R$:
\begin{equation}
\label{P=Pi}
P_R = \Pi\; ,
\end{equation}
\begin{equation}
\label{n_R=n_inf}
n_R = n_{\infty}\; .
\end{equation}
The subscripts $\infty$ denotes that the equilibrium concentration $n_\infty$ is related with the equilibrium near the flat surface of phase separation ($R \rightarrow \infty$). From Eq.~(\ref{P=Pi}) it follows that the gas concentration in the bubble $n_g$ is constant. Using the ideal gas law we have
\begin{equation}
\label{n_g}
n_g = \frac{\Pi}{k T}\; .
\end{equation}
When Eqs.~(\ref{R>>Laplace}), (\ref{P=Pi}), and (\ref{n_R=n_inf}) are fulfilled and, therefore, Laplace pressure is negligible, the bubble dynamics is evident for the case of steady-state diffusion and can be even described analytically for the non-steady case [\ref{Grinin-Kuni-Gor}, \ref{Scriven-1959}].

From the moment of bubble nucleation and till Eq.~(\ref{R>>Laplace}) becomes valid, Laplace pressure influences bubble growth. Thus both quantities $P_R$ and $n_R$ become radius-dependent (and, therefore, time-dependent). For $P_R$ now we have
\begin{equation}
\label{Mech-Eq}
P_R = \Pi + \frac{2 \sigma}{R}\; .
\end{equation}

In order to write the equation for $n_R$ we need to know the solubility law. For the simplest case of Henry's law the equilibrium concentration of dissolved gas is proportional to the corresponding pressure
\begin{equation}
\label{Chem-Eq-Henry}
\frac{n_R}{n_\infty} = \frac{P_R}{\Pi}\; .
\end{equation}
However, Henry's law is fulfilled only in such systems where there is no molecules dissociation during gas dissolution. In another important case when the gas molecules completely dissociate in the solvent into two parts the Henry's law is replaced with so-called Sievert's law (see e.~g. [\ref{Darken-Gurry}]), and, therefore, Eq.~(\ref{Chem-Eq-Henry}) is replaced with the following one:
\begin{equation}
\label{Chem-Eq}
\frac{n_R}{n_\infty} = \sqrt{\frac{P_R}{\Pi}}\; .
\end{equation}
Sievert's law is fulfilled for water vapor dissolved in a silicate melt [\ref{Stolper}]. Such solutions are important both for glass production [\ref{Cable-Frade}] and for volcanic systems [\ref{Navon}, \ref{Chernov-2004}].

The replacement of Eq.~(\ref{Chem-Eq-Henry}) with Eq.~(\ref{Chem-Eq}) means that the boundary condition for the gas diffusion problem will be different for Henry's and Sievert's laws. The change of boundary condition, as we will see further, leads to the change of bubble dynamics.

\section{Bubble Dynamics Equation}

After the nucleation of the bubble, when its growth can be considered regular (i.e. the bubble cannot be dissolved by fluctuations), its growth is governed by the diffusion flux of gas molecules into it. In this paper we will study the case when the diffusion flux can be assumed as steady. The conditions when such approximation is valid will be discussed further.

Taking into account the equality of gas concentration at the bubble surface to the equilibrium concentration $n_R$ and the equality of gas concentration far from the bubble to the initial concentration $n_0$, we can write a simple expression for the steady diffusion flux density $j_D$
\begin{equation}
\label{Flux}
j_D = D \frac{n_0 - n_R}{R}\; .
\end{equation}
Here $D$ is the diffusion coefficient of gas molecules in the pure solvent (we assume that the solution is diluted).

Now let us write the expression for the number of gas molecules $N$ in the bubble. Exploiting the ideal gas law and using Eq.~(\ref{Mech-Eq}) we have
\begin{equation}
\label{dN/dt-simple}
N = \frac{4 \pi}{3 k T} R^3 \left[ \Pi + \frac{2 \sigma}{R}\right]\; .
\end{equation}
Differentiating Eq.~(\ref{dN/dt-simple}), using Eq.~(\ref{n_g}), we obtain
\begin{equation}
\label{dN/dt}
\frac{dN}{dt} = 4 \pi n_g R^2 \frac{dR}{dt} \left[ 1 + \frac{R_\sigma}{R} \right]\; ,
\end{equation}
where
\begin{equation}
\label{R_s}
R_\sigma \equiv \frac{4}{3} \frac{\sigma}{\Pi}
\end{equation}
is the characteristic size of the bubble.

Material balance between the dissolved gas and the gas in the growing bubble gives us the following equality
\begin{equation}
\label{Balance}
\frac{dN}{dt} = 4 \pi R^2 j_D\; .
\end{equation}
Substituting expression for diffusion flux density Eq.~(\ref{Flux}) and the rate of change of the number of molecules Eq.~(\ref{dN/dt}) into material balance equation Eq.~(\ref{Balance}) we have
\begin{equation}
\label{Dyn-Eq-1}
n_g \frac{dR}{dt} \left[ 1 + \frac{R_\sigma}{R} \right] = D \frac{n_0 - n_R}{R}\; .
\end{equation}
Or, exploiting Eqs.~(\ref{s}) and (\ref{n_g}), we equivalently have
\begin{equation}
\label{Dyn-Eq-2}
R \dot{R} \left[ 1 + \frac{R_\sigma}{R} \right] = D s \frac{n_0 - n_R}{n_\infty}\; .
\end{equation}

Using Eqs.~(\ref{zeta}), (\ref{Mech-Eq}) and, correspondingly, (\ref{Chem-Eq-Henry}) and (\ref{Chem-Eq}), the fraction in the r. h. s. of Eq.~(\ref{Dyn-Eq-2}) can be expressed as
\begin{equation}
\label{Conc-Frac-1-Henry}
\frac{n_0 - n_R}{n_\infty} = \zeta - \frac{2 \sigma}{\Pi R}
\end{equation}
for Henry's law, and
\begin{equation}
\label{Conc-Frac-1}
\frac{n_0 - n_R}{n_\infty} = \zeta + 1 - \sqrt{1 + \frac{2 \sigma}{\Pi R}}
\end{equation}
for Sievert's law. After exploiting in Eqs.~(\ref{Conc-Frac-1-Henry}) or (\ref{Conc-Frac-1}) the definition of $R_\sigma$ (Eq.~(\ref{R_s})), we have, correspondingly,
\begin{equation}
\label{Conc-Frac-Henry}
\frac{n_0 - n_R}{n_\infty} = \zeta - \frac{3}{2} \frac{R_\sigma}{R}
\end{equation}
and
\begin{equation}
\label{Conc-Frac}
\frac{n_0 - n_R}{n_\infty} = \zeta + 1 - \sqrt{1 + \frac{3}{2} \frac{R_\sigma}{R}}\; .
\end{equation}
Eqs.~(\ref{Conc-Frac-Henry}) and (\ref{Conc-Frac}) allow us to rewrite the equation of bubble dynamics Eq.~(\ref{Dyn-Eq-2}) in the final form, namely
\begin{equation}
\label{Dyn-Eq-Henry}
R \dot{R} \left[ 1 + \frac{R_\sigma}{R} \right] = D s \left[ \zeta - \frac{3}{2} \frac{R_\sigma}{R} \right]
\end{equation}
for Henry's solubility law; and
\begin{equation}
\label{Dyn-Eq}
R \dot{R} \left[ 1 + \frac{R_\sigma}{R} \right] = D s \left[ \zeta + 1 - \sqrt{1 + \frac{3}{2} \frac{R_\sigma}{R}} \right]
\end{equation}
for Sievert's solubility law.

While Eq.~(\ref{Dyn-Eq-Henry}) was obtained as early as in 1950 in paper by Epstein and Plesset [\ref{Epstein-Plesset}], Eq.~(\ref{Dyn-Eq}) was not obtained previously. In paper by Cable and Frade it was presented only a special case of Eq.~(\ref{Dyn-Eq}), when $\zeta = -1$:
\begin{equation}
\label{Dyn-Eq-Dissolution}
R \dot{R} \left[ 1 + \frac{R_\sigma}{R} \right] = - D s \sqrt{1 + \frac{3}{2} \frac{R_\sigma}{R}}
\end{equation}
(Eq.~(34) in [\ref{Cable-Frade}]). Such value of supersaturation corresponds to the dissolution of gas bubble in the pure solvent. And even for this special case of Eq.~(\ref{Dyn-Eq}) analytical solution was not obtained. We present the analytical solution of Eq.~(\ref{Dyn-Eq-Dissolution}) in Appendix B.

\section{Critical Bubble and Initial Conditions for the Bubble  Growth}

The obtained equations for the bubble growth dynamics Eqs.~(\ref{Dyn-Eq-Henry}) and (\ref{Dyn-Eq}) can be applied to the cases of both homogeneous and heterogeneous nucleation. Below we will consider only the homogeneous nucleation case.

Homogeneous nucleation of a gas bubble in a supersaturated solution means fluctuational mechanism of its appearance and needs a significant pressure drop (in the order of $10^3$ times). It is these conditions of bubble nucleation that take place in magmatic melts during volcanic eruptions [\ref{Chernov-2004}]. Since we decided to describe only regular growth of a bubble, we need to exclude the very process of nucleation from our examination and consider a bubble only when it is already supercritical.

Under the notion of critical bubble we understand, as usually [\ref{Slezov-2004}, \ref{Skripov-1972}], such a bubble which radius $R_c$ corresponds to the extremum of work of bubble formation. Critical bubble is in mechanical equilibrium with solution at the initial pressure $P_0$ and in chemical equilibrium with solution with concentration $n_0$
\begin{equation}
\label{P_Rc}
P_{R_c} = P_0\; ,
\end{equation}
\begin{equation}
\label{n_Rc}
n_{R_c} = n_0\; .
\end{equation}
These two conditions together with the solubility law unambiguously define the value of $R_c$.
Condition (\ref{P_Rc}), using Eq.~(\ref{Mech-Eq}), evidently gives us
\begin{equation}
\label{R_c-def}
R_c = \frac{2 \sigma}{P_0 - \Pi}\; ,
\end{equation}
and condition (\ref{n_Rc}) allows us to connect value $P_0$ with the supersaturation $\zeta$.

Substituting Eqs.~(\ref{P_Rc}) and (\ref{n_Rc}) in Eq.~(\ref{Chem-Eq-Henry}) (Henry's law) and Eq.~(\ref{Chem-Eq}) (Sievert's law) we have correspondingly
\begin{equation}
\label{n_0/n_inf-Henry}
\frac{n_0}{n_\infty} = \frac{P_0}{\Pi}
\end{equation}
and
\begin{equation}
\label{n_0/n_inf}
\frac{n_0}{n_\infty} = \sqrt{\frac{P_0}{\Pi}}\; .
\end{equation}
For Henry's law we have evidently from Eq.~(\ref{n_0/n_inf-Henry}) using Eq.~(\ref{zeta})
\begin{equation}
\label{P_0-zeta-Henry}
P_0 = \Pi \left( \zeta + 1 \right).
\end{equation}
And, substituting Eq.~(\ref{P_0-zeta-Henry}) in Eq.~(\ref{R_c-def}), we have
\begin{equation}
\label{R_c-Henry}
R_c = \frac{2 \sigma}{\Pi \zeta}\; ,
\end{equation}
the well-known expression for the radius of a critical bubble (e.~g. [\ref{Slezov-2004}, \ref{Skripov-1972}]).

For Sievert's law, when Eq.~(\ref{n_0/n_inf-Henry}) is replaced by Eq.~(\ref{n_0/n_inf}), we have
\begin{equation}
\label{P_0}
P_0 = \Pi \left( \frac{n_0}{n_\infty} \right)^2,
\end{equation}
or, using Eq.~(\ref{zeta}),
\begin{equation}
\label{P_0-zeta}
P_0 = \Pi \left( \zeta + 1 \right)^2.
\end{equation}
Finally, substituting Eq.~(\ref{P_0-zeta}) in Eq.~(\ref{R_c-def}) for the critical bubble radius, we have
\begin{equation}
\label{R_c}
R_c = \frac{2 \sigma}{\Pi \left[ (\zeta + 1)^2 - 1 \right]}\; .
\end{equation}

Using Eq.~(\ref{R_s}) and Eq.~(\ref{R_c-Henry}) or Eq.~(\ref{R_c}) correspondingly we can write the relations between $R_\sigma$ and $R_c$ for both Henry's and Sievert's laws
\begin{equation}
\label{R_s-R_c-Henry}
R_\sigma = \frac{2}{3}\zeta R_c\; ,
\end{equation}
\begin{equation}
\label{R_s-R_c}
R_\sigma = \frac{2}{3} \left[ (\zeta + 1)^2 - 1 \right] R_c\; ,
\end{equation}
which will be exploited later.

Here it is important to make a remark regarding the quantity $s$, the gas solubility. Eqs.~(\ref{zeta}) and (\ref{s}) give us
\begin{equation}
\label{s(zeta)}
s = \frac{n_0 k T}{\Pi} \frac{1}{\zeta + 1}\; .
\end{equation}
If Henry's law is fulfilled, we have $\zeta + 1 = P_0/\Pi$ and, therefore,
\begin{equation}
\label{s-Henry}
s = \frac{n_0 k T}{P_0}\; ,
\end{equation}
$s$ is a tabular value defined only by the initial state of the solution (before the pressure drop).

Since we consider Sievert's law, we have (see Eq.~(\ref{P_0-zeta})) $\zeta + 1 = \sqrt{P_0/\Pi}$; thus instead of Eq.~(\ref{s-Henry}), we obtain
\begin{equation}
\label{s-Sievert}
s = \frac{n_0 k T}{\sqrt{P_0 \Pi}}\; .
\end{equation}
Here it is convenient to introduce constant $K$ as the coefficient of proportionality in Sievert's law: $K \equiv n_0/\sqrt{P_0}$. It allows us to rewrite Eq.~(\ref{s-Sievert}) in the following form:
\begin{equation}
\label{s-Sievert-K}
s = K \frac{kT}{\sqrt{\Pi}}\; .
\end{equation}
Eq.~(\ref{s-Sievert-K}) shows us that in the case of Sievert's law the solubility value $s$ depends on the final state of the solution. In the current paper a solubility value for Sievert's law is understood as gas solubility at the final pressure $\Pi$ i.e. after the pressure drop.

To deal with the dynamics equations (\ref{Dyn-Eq-Henry}) and (\ref{Dyn-Eq}) one needs to provide it with a reasonable initial condition, i.~e. to choose the value of constant $R_i$, initial radius of the bubble, in the following equality
\begin{equation}
\label{R(0)}
R(t)|_{t=0} = R_i\; .
\end{equation}
In papers [\ref{Epstein-Plesset}, \ref{Cable-Frade}] there was no special meaning assigned to the value of $R_i$: the reason of appearance of the bubble was totally excluded from discussion. Here we assume that the bubble nucleated fluctuationally, i.~e. crossed the barrier corresponding to radius $R_c$. It means that $R_i$ has to be not less than $R_c$, but evidently we cannot use the radius of a critical bubble $R_c$ as the initial value for the radius.

A bubble that nucleates fluctuationally in the solution is capable of regular growth if it passed and, moreover, moved away from the near-critical region where fluctuations are still strong enough. Thus we choose, following [\ref{Kuchma-Gor-Kuni}], $R_i = 2 R_c$, i.~e.
\begin{equation}
\label{2Rc}
R(t)|_{t=0} = 2 R_c\; .
\end{equation}
Such value guarantees the absence of fluctuations and, as we will see further, provides us with convenient expressions.

The necessary condition for the fluctuational nucleation of a bubble is a high pressure drop ($P_0/\Pi \sim 10^3$) and, consequently, high supersaturation $\zeta \sim 10^3$ for Henry's law and $\zeta \sim 30 \div 40$ for Sievert's law. Further we will use the following strong inequalities for the supersaturation:
\begin{equation}
\label{Strong-Zeta-Henry}
\zeta \gg 10
\end{equation}
for Henry's law, and
\begin{equation}
\label{Strong-Zeta}
\zeta \gg 1
\end{equation}
for Sievert's law.

\section{Three Stages of Bubble Growth}

Before we find the explicit solution of equations (\ref{Dyn-Eq-Henry}) and (\ref{Dyn-Eq}), let us qualitatively describe the change of the character of bubble growth process with the increase of its size. This will allow us to mark out the representative stages of growth and to determine corresponding characteristic bubble sizes. Duration of the consecutive stages and the character of bubble radius time dependence on each stage will be considered in the next section. It has to be noticed that the stages of our interest do not have anything in common with the stages of evolution of the whole ensemble of bubbles during the decomposition of liquid solution supersaturated with gas.

It will be more convenient to consider both cases of solubility laws separately.

\subsection{Henry's Law}

Let us rewrite Eq.~(\ref{Dyn-Eq-Henry}) in the equivalent form which will be more appropriate for its analysis:
\begin{equation}
\label{Growth-Rate-Henry}
\dot{R} = D s \zeta \left(1 - \frac{R_c}{R} \right) \frac{1}{R} \left( \frac{1}{1 + R_\sigma/R} \right),
\end{equation}
where we took into account Eqs.~(\ref{R_s}) and (\ref{R_c-Henry}).

Each of the three co-factors dependent on $R$ emphasized in the right hand side of Eq.~(\ref{Growth-Rate-Henry}) describes its physically different contribution to the dynamics of the supercritical bubble growth process. Co-factor $1 - R_c/R$, increasing with $R$, corresponds to the fast (the scale of change of $R$ is $R_c$) increase of the driving force of the process (the value $n_0 - n_R$) with the growth of $R$. Co-factor $1/R$, decreasing with the growth of $R$, describes, as it is seen from Eq.~(\ref{Flux}), the contribution related to the decrease of the gradient of solution concentration near the bubble surface, which decreases the bubble growth rate with the growth of $R$. Finally, co-factor $1/(1 + R_\sigma/R)$, which increases with the growth of $R$, takes into account the counteraction of Laplace pressure to the bubble growth, i.~e. the fact that the bubble growth is facilitated by the reduction of Laplace pressure with the growth of $R$, while other factors are equal (the scale of its change is $R_\sigma$). Notwithstanding the mentioned reduction of counteraction, the resulting contribution of the last two factors always leads to the deceleration of growth with the increase of bubble size.

From Eqs.~(\ref{R_s-R_c-Henry}) and (\ref{Strong-Zeta-Henry}) it can be seen that $R_\sigma \gg R_c$. Using this inequality, Eq.~(\ref{Growth-Rate-Henry}) and boundary condition (\ref{2Rc}), we have that the growth rate of the bubble radius in its dependence on this radius has to reach the maximum value achieved at certain bubble radius $R_m$ from the interval $2 R_c \leq R \leq R_\sigma$. Thus it is natural to consider the growth in the following interval of sizes
\begin{equation}
\label{1-st}
2 R_c \leq R \leq R_m
\end{equation}
as a first stage of bubble evolution, where the determining factor is the increase of the driving force of the growth. At this stage bubble growth goes with the increasing in time rate, reaching its maximum at $R_m$.

In order to obtain $R_m$ we will consider the rate of bubble growth as a function of its radius and differentiate both parts of Eq.~(\ref{Growth-Rate-Henry}) with respect to $R$:
\begin{equation}
\label{d-Growth-Rate-Henry}
\frac{d \dot{R}}{dR} = D s \zeta \frac{R_c \left(R_c + R_\sigma \right) - \left(R - R_c \right)^2}{R^2 \left(R + R_\sigma \right)^2}\; .
\end{equation}
The quantity $R_m$ is defined by the extremal condition
\begin{equation}
\label{Extremum}
\left. \frac{d \dot{R}}{dR} \right|_{R = R_m} = 0\; ,
\end{equation}
which using Eq.~(\ref{d-Growth-Rate-Henry}) leads to the following result
\begin{equation}
\label{R_m-Henry-1}
R_m = R_c + \left( R_c^2 + R_c R_\sigma \right)^{1/2}\; .
\end{equation}
Taking into account the strong inequality $R_\sigma \gg R_c$, we can simplify the obtained expression for $R_m$:
\begin{equation}
\label{R_m-Henry}
R_m \simeq \left( R_c R_\sigma \right)^{1/2}\; .
\end{equation}

The second stage of the process will be when the bubble growth occurs within the interval of sizes
\begin{equation}
\label{2-nd}
R_m \leq R \leq R_\sigma\; .
\end{equation}
During all this stage, as $R_m \simeq (R_c R_\sigma)^{1/2} \gg R_c$, it is already valid that $R \gg R_c$, thus the driving force $n_0 - n_R$ remains practically constant. As a result, bubble growth decelerates, although, as it was noted above, the counteraction of Laplace pressure to the growth gradually is attenuated.
The Laplace contribution $2 \sigma/R$ to the pressure inside the bubble decreases during the second stage by $(R_\sigma/R_c)^{1/2}$ times and at the completion of this stage becomes comparable with the external pressure $\Pi$ contribution (from Eq.~(\ref{R_s}) we have strict equality $2 \sigma/R_\sigma = 3 \Pi/2$). It is this physical condition that defines the completion of the second stage.

On the subsequent, third stage, which corresponds to the interval of sizes
\begin{equation}
\label{3-rd}
R \geq R_\sigma\; ,
\end{equation}
monotonous decelerated bubble growth continues. At the same time, the role of Laplace pressure continues to decrease gradually, and the pressure $P_R$ inside the bubble approaches to a constant value equal to the external pressure $\Pi$. As it will be shown further, this process is rather protracted, so the concluding phase of the third stage, when the pressure inside the bubble practically does not change and the application  of self-similar solution [\ref{Grinin-Kuni-Gor}, \ref{Scriven-1959}] is possible, comes only in the interval of sufficiently large sizes of the bubble, when the condition $R \gg R_\sigma$ is satisfied with a certain reserve.

\subsection{Sievert's Law}

\begin{figure}
\centering
\includegraphics[width=1.0\linewidth]{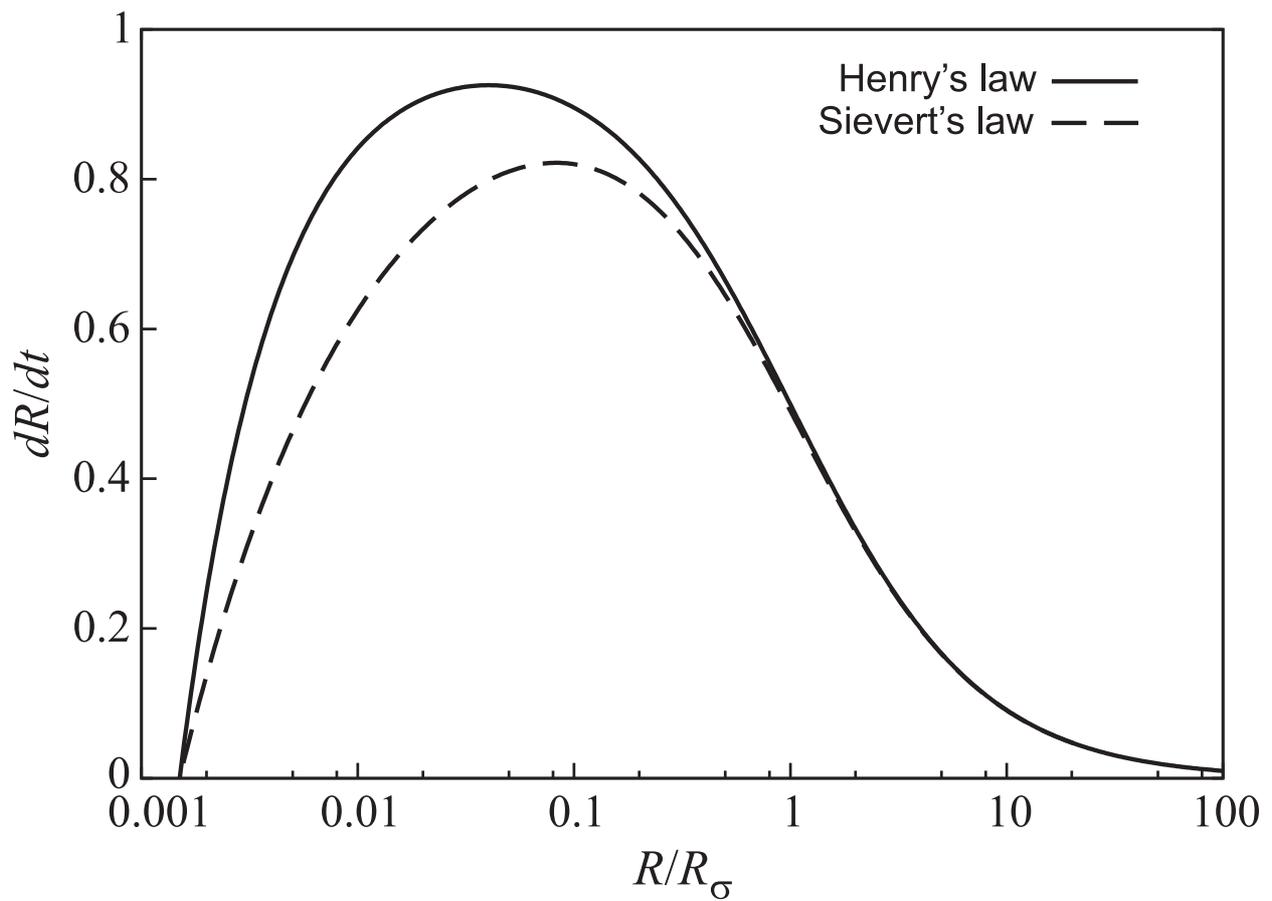}
\caption{Growth rate of bubble radius $dR/dt$ (measured in dimensionless units $D s \zeta/R_{\sigma}$) as a function of $R$. Solid curve corresponds to Henry's law, Eq.~(\ref{Growth-Rate-Henry}). Dashed curve corresponds to Sievert's law, Eq.~(\ref{Growth-Rate}). For both solubility laws $P_0/\Pi = 10^3$.}
\label{Fig:Rate}
\end{figure}

Let us investigate behavior of bubble growth rate $\dot{R}$ with the increase of bubble radius for Sievert's solubility law. From Eq.~(\ref{Dyn-Eq}) evidently stems.
\begin{equation}
\label{Growth-Rate}
\dot{R} = \frac{D s}{R + R_\sigma} \left[ \zeta + 1 - \sqrt{1 + \frac{3}{2} \frac{R_\sigma}{R}} \right]\; .
\end{equation}
Eq.~(\ref{Growth-Rate}) can be also written in the form similar to Eq.~(\ref{Growth-Rate-Henry}), to mark out three co-factors. The character of bubble growth rate is presented graphically in Fig. \ref{Fig:Rate}.

Differentiating Eq.~(\ref{Growth-Rate}) by $R$ we obtain
\begin{equation}
\label{d-Growth-Rate}
\frac{d\dot{R}}{dR} = \frac{D s}{(R + R_\sigma)^2} \left[ \frac{3}{4} \frac{R_\sigma (R + R_\sigma)}{\sqrt{1 + \frac{3}{2} \frac{R_\sigma}{R}} R^2} - \left( \zeta + 1 - \sqrt{1 + \frac{3}{2} \frac{R_\sigma}{R}} \right) \right]\; .
\end{equation}
It can be easily seen, that, as it is for Henry's law, here bubble radius growth rate as a function of variable $R$ also has the only maximum. Denote the corresponding radius as $R_m$, we can obtain its value from Eq.~(\ref{Extremum}). Using Eq.~(\ref{d-Growth-Rate}) we rewrite this equation as
\begin{equation}
\label{Extremum-Eq}
RR_\sigma + R_\sigma^2 + \frac{4}{3} R^2 \left( 1 + \frac{3}{2} \frac{R_\sigma}{R} \right) - \frac{4}{3} R^2 \sqrt{1 + \frac{3}{2} \frac{R_\sigma}{R}} (\zeta + 1) = 0\; .
\end{equation}

Assuming that the sought quantity $R_m$ is considerably less than $R_\sigma$, we will use for $R$ in Eq.~(\ref{Extremum-Eq}) the strong inequality $3 R_\sigma/ 2 R \gg 1$. Below we will need this inequality to be even more stronger
\begin{equation}
\label{Strong-For-Extr}
R_\sigma/ 3 R \gg 1\; .
\end{equation}
It will allow us to omit the second term in the brackets and the second addend in the square root in Eq.~(\ref{Extremum-Eq}). So this equation can be rewritten as:
\begin{equation}
\label{Extremum-Eq-1}
\sqrt{\frac{8}{3}} \sqrt{R_\sigma} (\zeta + 1) R^{3/2} - 3RR_\sigma - R_\sigma^2 = 0\; .
\end{equation}
Using Eq.~(\ref{Strong-For-Extr}) we can also omit the second addend in Eq.~(\ref{Extremum-Eq-1}) in comparison with the third one. After such a simplification Eq.~(\ref{Extremum-Eq-1}) becomes solvable; and for $R_m$ we have:
\begin{equation}
\label{R_m}
R_m \simeq \frac{\sqrt[3]{3}}{2} \frac{R_\sigma}{(\zeta + 1)^{2/3}}\; .
\end{equation}
Substituting Eq.~(\ref{R_m}) in the second addend in Eq.~(\ref{Extremum-Eq-1}) we can find a first-order correction to $R_m$ in Eq.~(\ref{R_m}). We have
\begin{equation}
\label{R_m_1}
R_m \simeq \frac{\sqrt[3]{3}}{2} \frac{R_\sigma}{(\zeta + 1)^{2/3}} \left( 1 + \frac{3 \sqrt[3]{3}}{2} \frac{1}{(\zeta + 1)^{2/3}}\right)^{2/3}\; .
\end{equation}
Numerical solution of Eq.~(\ref{Extremum-Eq}) for $\zeta = 30$ gives for inaccuracy of approximate solution Eq.~(\ref{R_m_1}) the value less than $2\%$, and this value evidently decreases with the increase of $\zeta$.

As soon as we explained the behavior of the value $\dot{R}$ and found the value of $R_m$ we can use the ideas proposed for Henry's law to determine the stages of bubble growth. On each stage the character of bubble growth is different from any other. The three consecutive stages of growth are still defined by Eqs.~(\ref{1-st}), (\ref{2-nd}) and (\ref{3-rd}).

In the next two sections we will obtain the time dependence of bubble radius $R$ for each stage and duration of each stage. Also we will obtain conditions of steadiness of bubble growth for each stage.

\section{Time Dependence of  Bubble Radius}

Let us now solve the differential equations (\ref{Dyn-Eq-Henry}) and (\ref{Dyn-Eq}) for the time dependence of the bubble radius with the initial condition Eq.~(\ref{R(0)}) of homogeneous nucleation of the bubble. As it was in the previous section it is more convenient here to consider Henry's and Sievert's laws separately.

\subsection{Henry's Law}

At first we will rewrite Eq.~(\ref{Dyn-Eq-Henry}) in the form which is appropriate for integration:
\begin{equation}
\label{For-Integration}
R \dot{R} + (R_\sigma + R_c) \dot{R} + (R_\sigma + R_c) R_c \frac{\dot{R}/R_c}{R/R_c - 1} = D s \zeta\; .
\end{equation}
Integrating Eq.~(\ref{For-Integration}), we obtain
\begin{equation}
\label{Integrated}
\frac{R^2}{2} + (R_\sigma + R_c) R + (R_\sigma + R_c) R_c \ln\left(\frac{R}{R_c} - 1\right) = D s \zeta \left(t + \tau \right)\; ,
\end{equation}
where $\tau$ is the constant which has time dimensionality and which is defined by the initial value of radius at time $t=0$. Using initial condition (\ref{R(0)}), from Eq.~(\ref{Integrated}) we find
\begin{equation}
\label{tau}
\tau = \frac{2 R_c (R_\sigma + 2 R_c)}{D s \zeta}\; .
\end{equation}
Excluding time $\tau$ from Eq.~(\ref{Integrated}) by means of Eq.~(\ref{tau}), we obtain
\begin{equation}
\label{Integrated-1}
\frac{R^2 - 4 R_c^2}{2} + (R_\sigma + R_c)(R_\sigma - 2 R_c) + (R_\sigma + R_c) R_c \ln\left(\frac{R}{R_c} - 1\right) = D s \zeta t\; .
\end{equation}

Validity of the general relation Eq.~(\ref{Integrated-1}), which strictly takes into account Laplace pressure influence on the bubble growth process, is limited only by the condition of applicability of steady approximation Eq.~(\ref{Flux}) for the diffusion flux. Equation (\ref{Integrated-1}) does not imply the smallness of quantity $R_c/R_\sigma$ which follows from Eqs.~(\ref{R_s-R_c-Henry}) and (\ref{Strong-Zeta-Henry}). Eq.~(\ref{Integrated-1}) complies with the results obtained in papers [\ref{Epstein-Plesset}, \ref{Cable-Frade}] for the particular case of steady growth of a bubble.

Now let us consider the third stage of bubble growth, when $R \geq R_\sigma$. First of all, let us note that at the end of the first stage, when the bubble radius $R$ approaches the value $(R_cR_\sigma)^{1/2}$, and with even more assurance on the second and the third stages, one can neglect the logarithmic addend in Eq.~(\ref{Integrated-1}). Moreover, since during the third stage the main contribution in the l. h. s. of Eq.~(\ref{Integrated-1}) gradually tends to the first addend, equal to $R^2/2$; and the contribution of the second addend (influence of Laplace pressure) decreases, neglect of logarithmic contribution becomes fairly justified. As a result,  Eq.~(\ref{Integrated-1}) conformably to the third stage of bubble growth can be written in the form of
\begin{equation}
\label{Growth-Quadratic}
\frac{R^2}{2} + R_\sigma R = D s \zeta t\; .
\end{equation}

\subsection{Sievert's Law}

Unlike the case of Henry's law (Eq.~(\ref{Dyn-Eq-Henry})) solution of Eq.~(\ref{Dyn-Eq}) in general case is too cumbersome and, as we will ensure in the current section, when inequality (\ref{Strong-Zeta}) is fulfilled, is even not necessary. Here we present solutions of this equation for two particular cases:
\begin{equation}
\label{Small-R}
R \ll R_\sigma
\end{equation}
-- the first and the second stages and
\begin{equation}
\label{Large-R}
R \gg R_c
\end{equation}
-- the second and the third stages. When strong inequality Eq.~(\ref{Strong-Zeta}) is fulfilled, these two cases cover the whole range $R \geq 2 R_c$ of regular growth of the bubble radius, and that is the reason why general solution of Eq.~(\ref{Dyn-Eq}) is not necessary for the system under consideration.

Let us begin with the case when inequality Eq.~(\ref{Small-R}) is fulfilled. In this case we can omit $1$ in comparison with $R_\sigma/R$ in the l. h. s. of Eq.~(\ref{Dyn-Eq}) and we can also omit $1$ in comparison with the fraction $3R_\sigma / 2R$ in the r. h. s. of this equation. Thus we have
\begin{equation}
\label{Growth-Small-R}
\dot{R} = \frac{D s}{R_\sigma} \left[ \zeta + 1 - \sqrt{\frac{3}{2} \frac{R_\sigma}{R}} \right]\; .
\end{equation}
Separating variables and exploiting Eq.~(\ref{R_s-R_c}) for $R_\sigma$ under the square root, we can rewrite Eq.~(\ref{Growth-Small-R}) in the form which allows its integration:
\begin{equation}
\label{Growth-Small-R-1}
\frac{dR}{1 - \sqrt{R_c/R}} = \frac{D s (\zeta + 1)}{R_\sigma} dt\; .
\end{equation}
Integrating Eq.~(\ref{Growth-Small-R-1}) with initial condition (\ref{R(0)}), we finally have
\begin{equation}
\label{Growth-Small-R-Final}
R - 2 R_c + R_c \ln \left( \frac{R}{R_c} - 1 \right) + 2 \sqrt{R_c} \left( \sqrt{R} - \sqrt{2 R_c} \right) 
 + R_c \ln \left(\frac{\sqrt{2} + 1}{\sqrt{2} - 1} \frac{\sqrt{R} - \sqrt{R_c}}{\sqrt{R} + \sqrt{R_c}} \right)
= \frac{D s}{R_\sigma}(\zeta + 1) t\; .
\end{equation}
This formula is different from Eq.~(\ref{Integrated-1}), which means that, when inequality (\ref{Small-R}) is fulfilled, there is a significant difference in the character of growth between Sievert's law and Henry's law.

Now let us proceed with the other case. At first, using Eq.~(\ref{R_s-R_c}) we can rewrite Eq.~(\ref{Dyn-Eq}) equivalently in the form of
\begin{equation}
\label{Growth-R_c}
\frac {\dot{R} \left[ R + R_\sigma \right]}{1 - \sqrt{\frac{1}{(\zeta + 1)^2} + \frac{R_c}{R}}} = D s (\zeta + 1)\; .
\end{equation}
This form makes it obvious that, when strong inequality (\ref{Large-R}) together with inequality (\ref{Strong-Zeta}) are fulfilled, the whole square root in the denominator of the l. h. s. of Eq.~(\ref{Growth-R_c}) can be omitted in comparison with $1$. Therefore, we have
\begin{equation}
\label{Growth-Large-R}
\dot{R} \left[ R + R_\sigma \right] = D s (\zeta + 1)\; .
\end{equation}
This expression can be easily integrated. But the use of the initial condition Eq.~(\ref{R(0)}) is not just as a result of the fulfillment of strong inequality (\ref{Large-R}). There is arbitrariness in the choice of the initial condition for integration of Eq.~(\ref{Growth-Large-R}), but the most convenient is to choose a condition at such an "average" radius which simultaneously satisfies Eqs.~(\ref{Small-R}) and (\ref{Large-R}), e.~g.
\begin{equation}
\label{Ra}
\left. R(t) \right|_{t=t_a} = R_a \equiv \sqrt{R_c R_{\sigma}}\; .
\end{equation}
Due to $R_a \ll R_{\sigma}$, we can use Eq.~(\ref{Growth-Small-R-Final}) to obtain the explicit value for time $t_a$ defined in Eq.~(\ref{Ra}). Using inequality $R_a \gg R_c$, Eq.~(\ref{Growth-Small-R-Final}) gives us
\begin{equation}
\label{ta}
t_a \simeq \frac{R_a R_{\sigma}}{D s (\zeta + 1)}\; .
\end{equation}
Now, integrating Eq.~(\ref{Growth-Large-R}) with initial condition (\ref{Ra}), we have
\begin{equation}
\label{Growth-Large-R-1}
\frac{R^2 - R_a^2}{2} + R_\sigma (R - R_a)= D s (\zeta + 1) (t - t_a)\; ,
\end{equation}
or, rewriting $R_a$ using Eq.~(\ref{Ra}) and $t_a$ using Eq.~(\ref{ta}), we have
\begin{equation}
\label{Growth-Large-R-2}
\frac{R^2}{2} + R_\sigma R - \frac{R_{\sigma}R_c}{2}= D s (\zeta + 1) t\; .
\end{equation}

\begin{figure}
\centering
\includegraphics[width=1.0\linewidth]{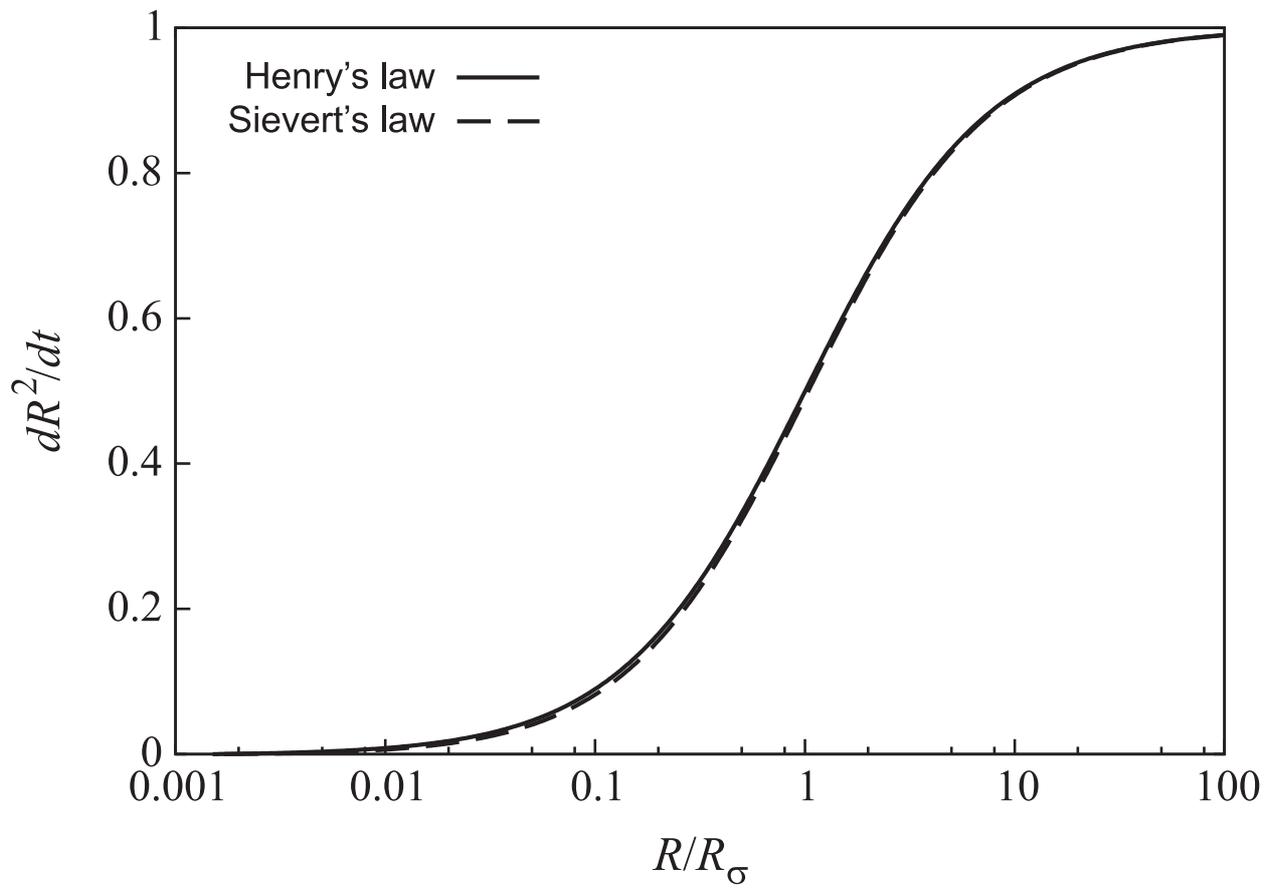}
\caption{Growth rate of bubble radius squared $dR^2/dt$ (measured in dimensionless units $2 D s \zeta$) as a function of $R$. Solid curve corresponds to Henry's law, Eq.~(\ref{Growth-Rate-Henry}). Dashed curve corresponds to Sievert's law, Eq.~(\ref{Growth-Rate}). For both solubility laws $P_0/\Pi = 10^3$.}
\label{Fig:Rate2}
\end{figure}

With the increase of $R$ the contribution of the third addend in the l. h. s. of Eq.~(\ref{Growth-Large-R-2}) decreases and at $R \geq R_{\sigma}$ (on the third stage) we can already write
\begin{equation}
\label{Growth-Large-R-Final}
\frac{R^2}{2} + R_\sigma R = D s (\zeta + 1) t\; .
\end{equation}
If we also used in Eq.~(\ref{Growth-Large-R-Final}) strong inequality (\ref{Strong-Zeta}), Eq.~(\ref{Growth-Large-R-Final}) will become identical to Eq.~(\ref{Growth-Quadratic}). It means that, when inequality $R \geq R_{\sigma}$ is fulfilled, any difference in the character of growth between Sievert's law and Henry's law disappears.

When the bubble radius becomes as large as
\begin{equation}
\label{Very-Large-R}
R \gg R_\sigma\; ,
\end{equation}
Eq.~(\ref{Growth-Quadratic}) transforms into the well-known Scriven's [\ref{Scriven-1959}] self-similar dependence for the steady-state case
\begin{equation}
\label{Growth-Large-R-ss}
R^2  = 2 D s \zeta t\; ,
\end{equation}
when $dR^2/dt = \mbox{const}$. This trend of Eq.~(\ref{Growth-Quadratic}) toward Eq.~(\ref{Growth-Large-R-ss}) was discussed in detail in [\ref{Kuchma-Gor-Kuni}]. This trend is presented graphically in Fig. \ref{Fig:Rate2}.

\section{Duration of the Consecutive Stages}

\subsection{Henry's Law}

Eq.~(\ref{Integrated-1}), giving the explicit dependence of bubble radius on time, allows us, in particular, to find characteristic times corresponding to consecutive stages. According to Eqs.~(\ref{1-st}) and (\ref{R_m-Henry-1}), at the end of the first stage, the bubble radius reaches the value $R_m = (R_c R_\sigma)^{1/2}$. Substituting value $R = R_m$ to Eq.~(\ref{Integrated-1}) and considering that (by virtue of inequalities $R_c \ll R_m \ll R_\sigma$) the main contribution to the l. h. s. of Eq.~(\ref{Integrated-1}) is made by the second addend, we obtain the expression for the first stage duration $t_1$
\begin{equation}
\label{t_1-Henry}
t_1 = \frac{R_\sigma^2}{Ds\zeta} \left(\frac{R_c}{R_\sigma}\right)^{1/2}\; .
\end{equation}
Using Eq.~(\ref{R_s-R_c-Henry}), expression (\ref{t_1-Henry}) can be also presented in the form
\begin{equation}
\label{t_1-z-Henry}
t_1 = \left(\frac{3}{2}\right)^{1/2} \frac{R_\sigma^2}{D s \zeta^{3/2}}\; .
\end{equation}
As it follows from Eq.~(\ref{t_1-z-Henry}), with the increase of initial supersaturation of the solution, the
first stage duration decreases proportionally to $1/\zeta^{3/2}$.

The second stage of bubble growth starts at the time point $t_1$ and finishes at the time point $t_2$ defined by the condition $\left. R \right|_{t = t_2} = R_\sigma$. Substituting value $R = R_\sigma$ into Eq.~(\ref{Integrated-1}) and considering that, by virtue of inequality $R_c \ll R_\sigma$, the main contribution to the l. h. s. of Eq.~(\ref{Integrated-1}) is made by the first and the second addends, we obtain the expression for $t_2$
\begin{equation}
\label{t_2-Henry}
t_2 = \frac{3}{2} \frac{R_\sigma^2}{D s \zeta}\; .
\end{equation}
As one can see from this expression, $t_2$ dependence on initial solution supersaturation is defined by multiplier $1/\zeta$. The duration of the second stage is much longer than the duration of the first stage, as from Eqs.~(\ref{t_1-z-Henry}) and (\ref{t_2-Henry}) it follows that
\begin{equation}
\label{t_2/t_1-Henry}
\frac{t_2}{t_1} = \left(\frac{3 \zeta}{2}\right)^{1/2} \gg 1\; .
\end{equation}

Now let us consider the third stage of bubble growth, when $R \geq R_\sigma$ and $t \geq t_2$. First of all, let us note that at the end of the first stage, when the bubble radius $R$ approaches the value $(R_c R_\sigma)^{1/2}$, and with even more assurance on the second and the third stages, one can neglect the logarithmic addend in Eq.~(\ref{Integrated-1}) (it was taken into account earlier, when Eqs.~(\ref{t_1-Henry}) and (\ref{t_2-Henry}) were obtained).

Due to Eq.~(\ref{3-rd}) the duration of the third stage of bubble growth is infinite. But it is reasonable to estimate the time from the beginning of the third stage and to the moment when bubble radius reaches the value $R_0$ defined in [\ref{Grinin-Kuni-Gor}],
\begin{equation}
\label{R_0}
R_0 \equiv 20 \times 2 \sigma / \Pi\; .
\end{equation}
It is assumed \textit{a priori} in [\ref{Grinin-Kuni-Gor}], that when radius reaches value $R_0$ the influence of Laplace pressure on the bubble growth is negligible.

As it was found in the previous section, on the third stage of bubble growth the time dependence of bubble radius is given by the simple equation, Eq.~(\ref{Growth-Quadratic}). Let us introduce time $t_3$ as a duration of bubble growth in the size interval $R_\sigma \leq R \leq R_0$. Using Eqs.~(\ref{Growth-Quadratic}) and (\ref{R_0}) we evidently have
\begin{equation}
\label{t_3-Henry}
t_3 \simeq 480 \frac{R_\sigma^2}{D s \zeta}\; .
\end{equation}
The duration of the third stage is much longer than the duration of the second stage (and, moreover, the first stage), as from Eqs.~(\ref{t_2-Henry}) and (\ref{t_3-Henry}) it follows that
\begin{equation}
\label{t_3/t_2-Henry}
\frac{t_3}{t_2} \simeq 320\; .
\end{equation}

Eq.~(\ref{t_3/t_2-Henry}) and strong inequality (\ref{t_2/t_1-Henry}) allow us to evaluate the whole duration of bubble growth in the interval of sizes $2 R_c \leq R \leq R_0$ as $t_3$. In [\ref{Grinin-Kuni-Gor}] the estimation of this duration $t_0$ was given by the following equation
\begin{equation}
\label{t_0}
t_0 \simeq \frac{R_0^2}{2 D s \zeta}\; .
\end{equation}
Using Eqs.~(\ref{R_s}) and (\ref{R_0}) we have $t_0 \simeq 450 \frac{R_\sigma^2}{D s \zeta}$; and therefore
\begin{equation}
\label{t_3-t_0}
\frac{t_3 - t_0}{t_3} \simeq 6\%\; .
\end{equation}

\subsection{Sievert's Law}

Using Eqs.~(\ref{Growth-R_c}) and (\ref{R_m_1}) we can obtain expressions for duration of the first two stages $t_1$ and $t_2$ for the Sievert's law analogous to Eqs.~(\ref{t_1-Henry}) and (\ref{t_2-Henry}). These expression are  cumbersome, while qualitatively the time-scale hierarchy for Sievert's law is similar to the one for Henry's law. Therefore we will not present these expressions here.

As long as on the third stage the dynamic equation of bubble growth for Henry's law Eq.~(\ref{Growth-Quadratic}) and for Sievert's law Eq.~(\ref{Growth-Large-R-Final}) are exactly the same (we need to account $\zeta \gg 1$ also), all the results presented above for the third stage duration are valid for Sievert's law also.

\section{Steady Flux Conditions}

The diffusion flux of molecules toward a growing bubble can be considered steady when the bubble growth is slow enough in comparison with the "diffusion cloud" growth. To be more exact, the radius of the bubble has to be much smaller than the radius of this cloud, the diffusion length. We can express it as
\begin{equation}
\label{R<<l}
R \ll (D t_R)^{1/2}\; ,
\end{equation}
where $t_R$ is the characteristic time of the bubble radius change, $t_{R} \equiv R/\dot{R}$ the time in which the bubble radius changes significantly. Evidently, Eq.~(\ref{R<<l}) can be rewritten as
\begin{equation}
\label{RR<<1}
\left( R \dot{R} / D \right)^{1/2} \ll 1\; .
\end{equation}
We can make this condition more explicit by means of Eq.~(\ref{Dyn-Eq-Henry}) and Eq.~(\ref{Dyn-Eq}) that give us correspondingly:
\begin{equation}
\label{Steady-General-Henry}
\left( s \zeta \frac{R-R_c}{R + R_\sigma} \right)^{1/2} \ll 1
\end{equation}
for Henry's law, and
\begin{equation}
\label{Steady-General}
\left( s \frac{(\zeta + 1) - \sqrt{1 + \frac{3}{2} \frac{R_\sigma}{R}}}{1 + \frac{R_\sigma}{R}} \right)^{1/2} \ll 1
\end{equation}
for Sievert's law.

Now exploiting Eqs.~(\ref{Steady-General-Henry}) and (\ref{Steady-General}) let us obtain the conditions for diffusion flux to be steady on each stage defined above. We will write these conditions as inequalities for the value of solubility, not for supersaturation. The value of supersaturation is already fixed by Eqs.~(\ref{Strong-Zeta-Henry}) and (\ref{Strong-Zeta}).

\begin{figure}\centering
\includegraphics[width=1.0\linewidth]{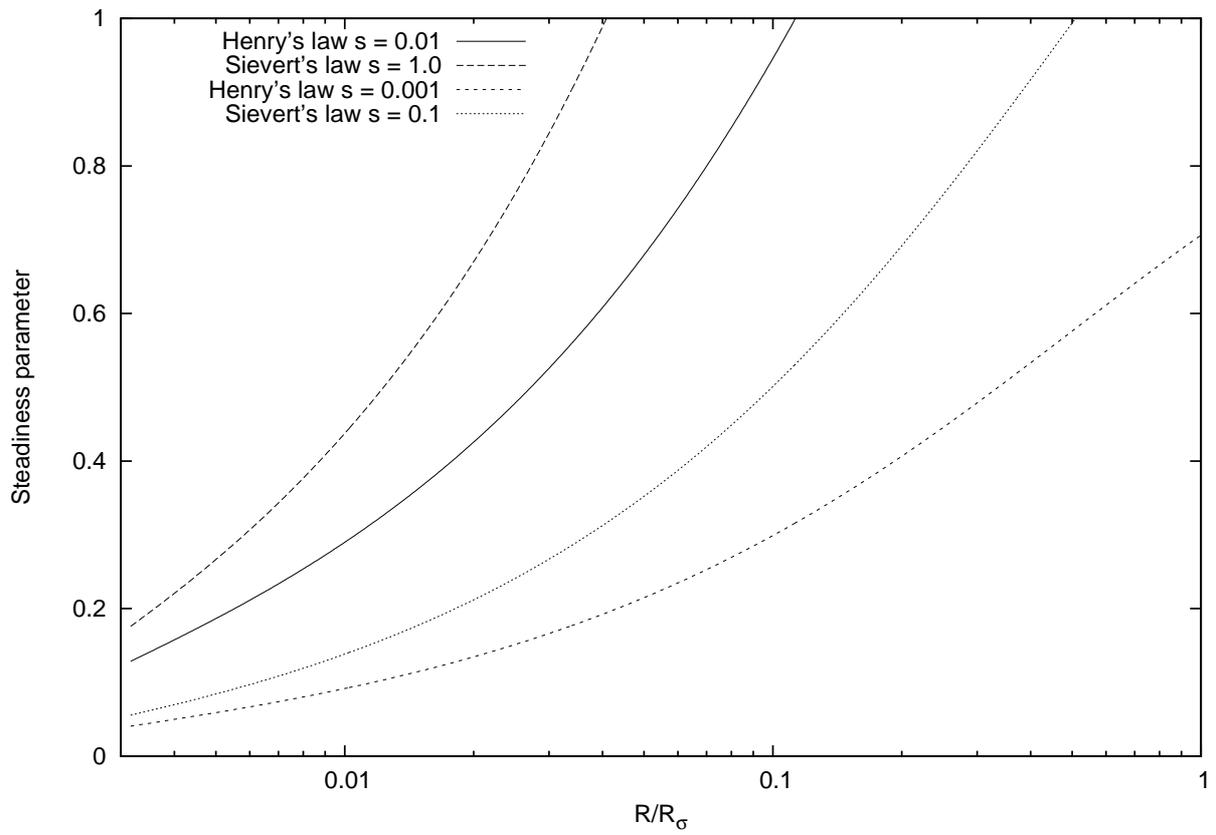}
\caption{Parameter characterizing the steadiness of bubble growth as a function of bubble radius $R$, l. h. s. of Eq.~(\ref{RR<<1}) at $P_0/\Pi = 10^3$. Curves are representing (from upper to lower) Sievert's law at $s = 1$, Henry's law at $s = 10^{-2}$, Sievert's law at $s = 10^{-1}$ and Henry's law at $s = 10^{-3}$.}
\label{Fig:Steadiness}
\end{figure}

Obviously the larger the bubble is the more strict condition for steadiness is (see Fig. \ref{Fig:Steadiness}). The general condition for bubble growth to be steady at any time is the condition at $R \gg R_\sigma$. From both Eqs.~(\ref{Steady-General-Henry}) and (\ref{Steady-General}) we have
\begin{equation}
\label{Steady-3}
s^{1/2} \ll \left( \frac{1}{\zeta} \right)^{1/2}\; .
\end{equation}
This condition is sufficient for steadiness on the third stage for both Henry's and Sievert's laws.

For bubble growth to be steady during the whole second stage it is sufficient to be steady at $R = R_\sigma$. This condition,
using Eq.~(\ref{Strong-Zeta-Henry}), transforms Eq.~(\ref{Steady-General-Henry})
\begin{equation}
\label{Steady-2}
s^{1/2} \ll \left( \frac{2}{\zeta} \right)^{1/2}\; .
\end{equation}
Exactly the same result will be for Sievert's law (see Eq.~(\ref{Strong-Zeta}) and Eq.~(\ref{Steady-General}) at $R = R_\sigma$).

For bubble growth to be steady during the whole first stage it is sufficient to be steady at $R = R_m$. It is to be reminded that the quantity $R_m$ has different values for Henry's and Sievert's laws (see Eq.~(\ref{R_m-Henry}) and Eq.~(\ref{R_m})). For Henry's law, using Eq.~(\ref{R_s-R_c}) this condition leads to
\begin{equation}
\label{Steady-1-Henry}
s^{1/2} \ll \left( \frac{2}{3 \zeta} \right)^{1/4}\; .
\end{equation}
And for Sievert's law to
\begin{equation}
\label{Steady-1}
s^{1/2} \ll \left( \frac{2}{3 \zeta} \right)^{1/6}\; .
\end{equation}
Obtaining Eq.~(\ref{Steady-1}) we used not only Eq.~(\ref{Strong-Zeta}), but even more strict condition $\zeta^{2/3} \gg 1$. This condition is still fulfilled than $\zeta \sim 30 \div 40$.

Finally, let us also write the condition of steadiness at the very beginning of the regular growth of the bubble, exploiting Eqs.~(\ref{Steady-General-Henry}) and (\ref{Steady-General}) with $R = 2 R_c$. For Henry's law we have
\begin{equation}
\label{Steady-0-Henry}
s^{1/2} \ll \left(\frac{2}{3}\right)^{1/2}\; ,
\end{equation}
and for Sievert's law
\begin{equation}
\label{Steady-0}
s^{1/2} \ll \zeta^{1/2}\; .
\end{equation}
While deriving Eq.~(\ref{Steady-0}), the numerical coefficient $\left( 3 \left( 1 - 1/\sqrt{2} \right) \right)^{1/2}$ in its l. h. s. was replaced with $1$ for shortness.

Let us mention the following interesting observation. Since for $R \gg R_{\sigma}$, the bubble dynamics is exactly the same for both Henry's and Sievert's solubility laws, the condition of steady growth for $R \gg R_{\sigma}$ is also the same (see Eq.~(\ref{Steady-3}) above). For the case of homogeneous nucleation, when the pressure drop $P_0/\Pi \sim 10^3$, the steady condition at $R \gg R_{\sigma}$ as a rule is violated in cases of both Henry's and Sievert's laws. For Henry's law, when $s \sim 10^{-2}$, it is violated due to high supersaturation values $\zeta \sim 10^{3}$. For Sievert's law, when corresponding supersaturation values are significantly less $\zeta \sim 30 \div 40$, values of gas solubility are significantly higher than for Henry's law (see next section), and that is the reason of the violation of the steady condition.
While for Henry's law solubility $s$ is a tabular value (see Eq.~(\ref{s-Henry})), for Sievert's law it can be adjusted via settlement of the final pressure $\Pi$ value. It is evident that condition (\ref{Steady-3}) can be satisfied at the given solubility value when one decreases the solution supersaturation $\zeta$ (the case of heterogeneous nucleation).

In order to satisfy the condition (\ref{Steady-3}) of steady growth the value of gas solubility $s$ has to be decreased. From Eq.~(\ref{P_0-zeta}) for the supersaturation we have
\begin{equation}
\label{zeta-Sievert}
\zeta = \sqrt{\frac{P_0}{\Pi}} - 1\; .
\end{equation}
Than, using Eqs.~(\ref{s-Sievert-K}), (\ref{zeta-Sievert}), let us rewrite product $s \zeta$ in the following form
\begin{equation}
\label{s-zeta}
s \zeta = K \frac{kT}{\sqrt{\Pi}} \left( \sqrt{\frac{P_0}{\Pi}} - 1 \right)\; .
\end{equation}
Eq.~(\ref{s-zeta}) shows us that in order to weaken limitation Eq.~(\ref{Steady-3}) one needs to increase the final pressure $\Pi$, leaving the ratio $P_0/\Pi$ constant.

\section{Bubble Growth in Volcanic Systems}

This section contains the analysis of the steady growth condition obtained in the previous section for bubble nucleation in volcanic systems. Previously, e. g. in papers [\ref{Chernov-2004}] and [\ref{Navon}]  the study of such systems exploited steady approximation without the analysis on its applicability.

For our evaluations we will use parameters from [\ref{Chernov-2004}] for the case of homogeneous nucleation of water vapor bubbles in a magmatic melt. We have
\begin{equation}
\label{data}
P_0 = 100~\mathrm{MPa} ~~~ \Pi = 0.1~\mathrm{MPa} ~~~ T = 1150~\mathrm{K} ~~~ w = 3\% ~~~ \rho_m = 2300~\mathrm{kg}/\mathrm{m}^3.
\end{equation}
Here $w$ is gas mass fraction of the dissolved gas (water vapor), and  $\rho_m$ is magma density.

Let us express the values of $s$ and $\zeta$ using data given. From Eq.~(\ref{zeta-Sievert}), in accordance with Eq.~(\ref{data}), we have $\zeta \simeq 31$. Then we need to calculate $n_0$ and substitute it into Eq.~(\ref{s-Sievert}) to obtain the value of solubility. As long as we are given mass density of magma and mass fraction of gas, it is convenient to write
\begin{equation}
\label{n_0-rho_0}
n_0 = \rho_0 \frac{N_A}{\mu}\; ,
\end{equation}
where $\rho_0$ is mass density of the dissolved gas, $N_A = 6 \times 10^{23}~\mathrm{mol}^{-1}$ is the Avogadro constant and $\mu = 1.8 \times 10^{-3}~\mathrm{kg}/\mathrm{mol}$ is the molar mass of the dissolved gas (water). Finally, we need to express the mass density of the dissolved gas. Evidently, we have
\begin{equation}
\label{rho_0}
\rho_0 = w \rho_m\; ,
\end{equation}
and, therefore,
\begin{equation}
\label{s-Chernov}
s = w \rho_m \frac{N_A k T}{\mu \sqrt{P_0 \Pi}}\; .
\end{equation}
Using data given by Wq.~(\ref{data}) in Eq.~(\ref{s-Chernov}), we have $s \simeq 12$.

Now we can see that for volcanic systems, where the pressure drop is of the order of $10^3$ and solubility is more than $1$, both conditions (\ref{Steady-3}) and (\ref{Steady-2}) are violated, and steady approximation is not valid for radii of the order of $R_{\sigma}$. Even in the very beginning of bubble regular growth, when $R = 2 R_c$, the steady condition (\ref{Steady-0}) is fulfilled only at its breaking point: the values of $\zeta^{1/2}$ exceed the value of $s^{1/2}$, but these values are of the same order of magnitude.

\section{Conclusions}

In the presented paper we obtained the equations for the bubble growth dynamics in the gas solution with Henry's and Sievert's solubility laws. We solved these equations analytically for case of bubble growth in strongly supersaturated solution. The equation for the Sievert's law was solved also for the case of bubble dissolution in the pure solvent.

We showed that, irrespective of the to gas solubility law, three characteristic stages could be marked out in the growth dynamics. During the first stage the bubble radius is growing with an increasing rate. On the second stage the growth rate decreases. The third stage, when the growth rate continues to decrease, begins when the Laplace pressure inside the bubble becomes comparable with the external pressure of the solution. We demonstrated that during the first two stages the time dependence of the bubble radius is different for the cases of Henry's and Sievert's laws, while during the third stage this distinction is no longer observed.
For both Henry's and Sievert's laws we obtained intervals within which the bubble radius changes on each stage, as well as durations of consecutive stages.

While obtaining the dynamic equations we assumed the diffusion flux to be steady. We obtained conditions when this steady approximation is applicable. We showed that usually, as the radius of the bubble increases, the steady regime of bubble growth gradually gives way to the nonsteady one.
Application of the obtained conditions for the volcanic system consisting of water vapor dissolved in the silicate melt showed that the process in such system as a rule cannot be considered as steady.

\section*{Appendix A: Effect of solvent viscosity on the bubble growth}

In the current paper we neglected the solvent viscosity. Its influence on bubble dynamics can be estimated using Rayleigh-Plesset equation (see e.g. [\ref{Brennen-1995}]). To take the solvent viscosity into account one needs to replace Eq.~(\ref{Mech-Eq}) with the following equality:
\begin{equation}
\label{Viscosity-1}
P_R = \Pi + \frac{2 \sigma}{R} + 4 \eta \frac{\dot{R}}{R}\; ,
\end{equation}
where $\eta$ is the dynamic viscosity of the solvent. The inertial terms in Rayleigh-Plesset equation are negligible for any reasonable bubble growth rate.

To neglect the viscous term in Eq.~(\ref{Viscosity-1}) (the third addend) in comparison with the surface tension term (the second addend) the following strong inequality has to be fulfilled:
\begin{equation}
\label{Viscosity-2}
\eta \ll \frac{\sigma}{2 \dot{R}}\; .
\end{equation}
This inequality, evidently, is equivalent to inequality $Pe \gg 1$, where $Pe$ is Peclet number.

The higher the bubble growth rate $\dot{R}$ is, the stronger the inequality (\ref{Viscosity-2}) is. The strongest condition takes place when $R = R_m$. Using Eqs.~(\ref{R_m-Henry}) and (\ref{R_m}) in Eqs.~(\ref{Growth-Rate-Henry}) and (\ref{Growth-Rate}) correspondingly, for both Henry's and Sievert's laws we have
\begin{equation}
\label{Viscosity-3}
\dot{R} \leq \left. \dot{R} \right|_{R = R_m} \simeq \frac{D s \zeta}{R_\sigma}\; .
\end{equation}
Therefore, using Eq.~(\ref{R_s}), we can rewrite strong inequality (\ref{Viscosity-2}) as
\begin{equation}
\label{Viscosity-4}
\eta \ll \frac{\sigma^2}{D s \zeta \Pi}\; ,
\end{equation}
where multiplier $2/3$ in the r. h. s. is omitted for shortness.

Let us estimate the value in the r. h. s. of inequality (\ref{Viscosity-4}). Typical values of surface tension both for water and for volcanic systems [\ref{Navon}] are $\rm \sigma \sim 10^{-1}~Nm^{-1}$; both for Henry's and Sievert's laws $s \zeta \sim 10$; diffusion coefficient $D \sim 10^{-11}~\rm m^2s^{-1}$ [\ref{Chernov-2004}]; pressure $\Pi \sim 10^5~\rm Pa$. Substituting these values, we have:
\begin{equation}
\label{Viscosity-5}
\eta \ll 10^{3}~\rm Pa~s\; .
\end{equation}
It should be noted that "common" liquids at normal conditions always satisfy this condition: for water we have $\eta \sim 10^{-3}~\rm Pa~s$ and even for glycerol $\eta \sim 1~\rm Pa~s$ [\ref{Landau-VI}].
For volcanic systems the values of viscosity that satisfy strong inequality (\ref{Viscosity-5}) are quite typical when $\rm SiO_2$ content is not too high (basalt, andesite and dactite melts) [\ref{Sparks-1978}]. However, for rhyolite melts ($\sim 70\%~\rm SiO_2$) viscosity can reach the values of $10^7~Pa~s$ [\ref{Navon}]; and, therefore, effect of solvent viscosity has to be taken into account.

We do not discuss here the oscillating settlement of the mechanical equilibrium between the bubble and the solution: as it was shown in [\ref{Trofimov-Melikhov-1994}] this settlement occurs much faster than the settlement of chemical equilibrium, unless the liquid viscosity is extremely low.

\section*{Appendix B: Dissolution of the gas bubble in a pure solvent: Sievert's solubility law}

Eq.~(\ref{Dyn-Eq-Dissolution}) allows us to obtain the radius-time relation for the bubble of arbitrary initial size $\left. R  \right|_{t=0} = R_i$ which is put in the pure solvent and also the time of its dissolution $t_d$, $\left. R  \right|_{t=t_d} = 0$.
Separating variables, we can rewrite Eq.~(\ref{Dyn-Eq-Dissolution}) as
\begin{equation}
\label{Dissolution-1}
\frac{1 + \frac{R_\sigma}{R}}{\sqrt{1 + \frac{3}{2} \frac{R_\sigma}{R}}} R dR = - D s dt\; ,
\end{equation}
or equivalently as
\begin{equation}
\label{Dissolution-2}
\left[ \sqrt{1 + \frac{3}{2} \frac{R_\sigma}{R}} - \frac{1}{2} \frac{\frac{R_\sigma}{R}}{\sqrt{1 + \frac{3}{2} \frac{R_\sigma}{R}}} \right] R dR = - D s dt\; .
\end{equation}
Using the variable $x \equiv \sqrt{1 + \frac{3}{2} \frac{R_\sigma}{R}}$ for integrating in Eq.~(\ref{Dissolution-2}) instead of $R$ we have
\begin{equation}
\label{Dissolution-3}
\frac{3}{2} R_\sigma^2 \left[ 3 \frac{x^2}{(x^2 - 1)^3} - \frac{1}{(x^2 - 1)^2} \right] dx = D s dt\; .
\end{equation}
Eq.~(\ref{Dissolution-3}) can be easily integrated and therefore we obtain
\begin{equation}
\label{Dissolution-4}
 \left[ \frac{1}{2} R \sqrt{R^2 + \frac{3}{2} R_\sigma R} - \frac{1}{8} R_\sigma \sqrt{R^2 + \frac{3}{2} R_\sigma R} \right. 
\left. \left. - \frac{3}{32} R_\sigma^2 \ln \left( \frac{4}{3} \frac{R}{R_\sigma} - \frac{4}{3} \frac{1}{R_\sigma}\sqrt{R^2 + \frac{3}{2} R_\sigma R} + 1 \right) \right] \right|^{R_i}_{R} = D s t\; .
\end{equation}
Eq.~(\ref{Dissolution-4}) gives us an explicit relation between the bubble radius $R$ and time $t$.

Let us find the time $t_d$ of total dissolution ($R \rightarrow 0$) from arbitrary initial radius $R_i$
\begin{equation}
\label{Dissolution-5}
 t_d = \frac{1}{D s} \left[ \frac{1}{2} R_i \sqrt{R_i^2 + \frac{3}{2} R_\sigma R_i} - \frac{1}{8} R_\sigma \sqrt{R_i^2 + \frac{3}{2} R_\sigma R_i} \right. 
\left. - \frac{3}{32} R_\sigma^2 \ln \left( \frac{4}{3} \frac{R_i}{R_\sigma} - \frac{4}{3} \frac{1}{R_\sigma}\sqrt{R_i^2 + \frac{3}{2} R_\sigma R_i} + 1 \right) \right]\; .
\end{equation}
The latter expression can be simplified for the two particular cases and short analytical expressions for time $t_d$ can be obtained. The first case is $R_i \gg R_\sigma$:
\begin{equation}
\label{Dissolution-6}
t_d = \frac{R_i^2}{2 D s}\; .
\end{equation}
And the second is $R_i \ll R_\sigma$:
\begin{equation}
\label{Dissolution-7}
t_d = \left( \frac{2}{3} \right)^{3/2} \frac{R_i^{3/2} R_\sigma^{1/2}}{D s}\; .
\end{equation}
Let us also obtain the expression for dissolution time from $R_i = R_\sigma$
\begin{equation}
\label{Dissolution-8}
t_d = 0.435 \frac{R_\sigma^2}{Ds}\; .
\end{equation}

\section*{References}

{\small
\begin{enumerate}

\item \label{Sparks-1978}
R.~S.~J. Sparks,
 J. Volcanol. Geoth. Res. {\bf 3}, 1 (1978).

\item \label{Sahagian-1999}
D.~Sahagian,
 Nature {\bf 402}, 589 (1999).

\item \label{Chernov-2004}
A.~A. Chernov, V.~K. Kedrinskii, and M.~N. Davidov,
 J. Appl. Mech. Tech. Phys. {\bf 45}, 281 (2004).

\item \label{Slezov-2004}
V.~V. Slezov, A.~S. Abyzov, and Z.~V. Slezova,
 Colloid J. {\bf 66}, 575 (2004).

\item \label{Slezov-2005}
V.~V. Slezov, A.~S. Abyzov, and Z.~V. Slezova,
 Colloid J. {\bf 67}, 85 (2005).

\item \label{Navon}
O.~Navon, A.~Chekhmir, and V.~Lyakhovsky,
 Earth Planet. Sci. Lett. {\bf 160}, 763 (1998).

\item \label{Lensky}
N.~G. Lensky, R.~W. Niebo, J.~R. Holloway, V.~Lyakhovsky, and O.~Navon,
 Earth Planet. Sci. Lett. {\bf 245}, 278 (2006).

\item \label{Kuchma-Gor-Kuni}
A.~E. Kuchma, G.~Y. Gor, and F.~M. Kuni,
 Colloid J. {\bf 71}, 520 (2009).

\item \label{Gor-Kuchma-Sievert}
G.~Y. Gor and A.~E. Kuchma,
 J. Chem. Phys. {\bf 131}, 034507 (2009).

\item \label{Epstein-Plesset}
P.~S. Epstein and M.~S. Plesset,
 J. Chem. Phys. {\bf 18}, 1505 (1950).

\item \label{Stolper}
E.~Stolper,
 Contrib. Mineral. Petrol. {\bf 81}, 1 (1982).

\item \label{Cable-Frade}
M.~Cable and J.~R. Frade,
 Proc. R. Soc. London, Ser. A {\bf 420}, 247 (1988).

\item \label{Grinin-Kuni-Gor}
A.~P. Grinin, F.~M. Kuni, and G.~Y. Gor,
 Colloid J. {\bf 71}, 46 (2009).

\item \label{Scriven-1959}
L.~E. Scriven,
 Chem. Eng. Sci. {\bf 10}, 1 (1959).

\item \label{Darken-Gurry}
L.~S. Darken and R.~W. Gurry,
 {\em Physical Chemistry of Metals}
 (McGraw-Hill, New York, 1953).

\item \label{Skripov-1972}
V.~P. Skripov,
 {\em Metastable Liquids}
 (John Wiley and Sons, New York, 1974).

\item \label{Brennen-1995}
C.~E. Brennen,
 {\em Cavitation and Bubble Dynamics},
 Oxford University Press, New York, 1995.

\item \label{Landau-VI}
L.~D. Landau and E.~M. Lifshitz,
 {\em Fluid Mechanics }
 ((2nd ed.), Pergamon Press, Oxford, 1987).

\item \label{Trofimov-Melikhov-1994}
Y.~V. Trofimov, A.~A. Melikhov, and F.~M. Kuni,
 Colloid J. {\bf 56}, 246 (1994).

\end{enumerate}
}

\end{document}